\documentclass[twocolumn,amsmath,amssymb,floatfix,prl,preprintnumbers,footinbib, showpacs]{revtex4-1}
\newlength{\stdwidth}
\usepackage[dvips]{graphicx}

\usepackage{bbm}
\usepackage{amsfonts}
\usepackage{amsmath}
\usepackage{graphicx}
\usepackage[utf8]{inputenc}

\usepackage[usenames,dvipsnames]{xcolor}
\usepackage[colorlinks=true, linkcolor=RoyalBlue, citecolor=RoyalBlue]{hyperref}

\newcommand{\vsubref}[2]{\ref{#1}{\color{RoyalBlue}({#2})}}
\newcommand{\vsub}[1]{{#1}}

\newcommand{\subref}[2]{\ref{#1}{\color{RoyalBlue}({#2})}}
\newcommand{\sub}[1]{{#1}}

\newcommand{\eref}[1]{(\ref{#1})}

\usepackage[normalem]{ulem}
\usepackage{color}

\begin{document}
\title{
Using Topological Insulator Proximity to Generate Perfectly Conducting Channels \\
in Materials without Topological Protection
}

\author{Sven Essert}
\affiliation{Institut f\"ur Theoretische Physik, Universit\"at Regensburg, D-93040 Regensburg, Germany}

\author{Viktor Krueckl}
\affiliation{Institut f\"ur Theoretische Physik, Universit\"at Regensburg, D-93040 Regensburg, Germany}

\author{Klaus Richter}
\affiliation{Institut f\"ur Theoretische Physik, Universit\"at Regensburg, D-93040 Regensburg, Germany}

\date{\today}

\begin{abstract}
We show that hybrid structures of topological insulators and materials without
topological protection can be employed to create
perfectly conducting channels hosted in the non-topological part.
These states inherit the topological protection from the proximity of the  topological
insulator but are more fragile to time-reversal symmetry breaking because of their
extended character.
We explore their formation in the band structure of model hybrid systems as well as
realistic heterostructures involving HgTe/CdTe-based two-dimensional topological insulators.
Using numerical quantum transport calculations for the HgTe/CdTe material system 
we propose two experimental settings which allow for the detection of the induced
perfectly conducting channels, both in the localized and diffusive regime, by means of
magneto conductance and shot noise. 
\end{abstract}


\maketitle

One important feature of the topological classification of insulators~\cite{Schnyder2008} is the existence of gapless
states at interfaces between materials which differ in their topological quantum number.
In this way, one can understand the formation of quasi one-dimensional (1d) edge channels in quantum Hall systems and
in two-dimensional (2d) topological insulators (TI) at the interface between the topologically non-trivial
insulator and the trivial insulating vacuum.
These edge channels carry the unique property of being perfectly conducting,
even in presence of impurity scattering, corresponding to transmission eigenvalues of
one.
Such quantized, spin-polarized channels along the boundaries of a 2d TI crystal have already
been detected experimentally~\cite{Brune2012} and give rise to peculiar conductance phenomena such 
as the quantum spin Hall effect~\cite{V_Bernevig2006,Konig2007}.

\begin{figure}[t]
\centering
\includegraphics[width=\stdwidth]{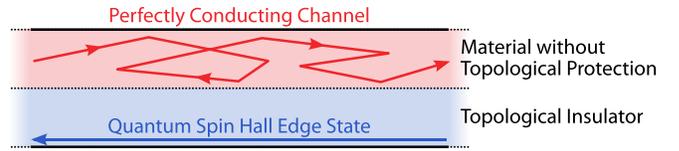}
\caption{ (\emph{Color online})
Sketch of a non-topological strip in proximity to a topological insulator inducing a
perfectly conducting channel in the normal material.
}
\label{figSketch}
\end{figure}

In this paper we show that perfectly conducting channels (PCCs) can be induced 
in materials {\em without} topological protection by the proximity of a 2d TI as sketched in
Fig.~\ref{figSketch}.
In such a hybrid structure the non-topological material could be a (disordered) metal, a gated
semi-conductor or even a conventional Anderson insulator.
Such a setup exploits the hybridization of the upper TI edge state with the bulk states
of the normal material, leading to an imbalance of left and right moving 
states in the normal material and thereby to a PCC therein~\cite{note1}.
The emergence of PCCs in isolated non-topological systems has been earlier predicted for carbon 
nanotubes~\cite{V_Ando2002} and graphene nanoribbons~\cite{V_Wakabayashi2007,V_Wurm2012}.
There the PCC was shown~\cite{V_Wakabayashi2007} to arise from an uneven separation of
left and right moving states associated with specific (valley) symmetries~\cite{V_Wurm2012}.
However, since atomic defects or short-ranged impurity potentials break this symmetry,
PCCs are rather fragile in these graphene-based systems and have not yet been observed therein.
Contrarily, the proximity-induced PCCs reported here are robust against any kind of disorder 
as long as time-reversal symmetry is preserved, even in presence of spin-orbit coupling.

To our knowledge this phenomenon has not been discussed before in 2d heterostructures, only 
layers made of 2d TIs nested in 3d systems~\cite{Hutasoit2011,Wang2013,Yang2013} and pure insulator heterostructures~\cite{xiao2013proximity}
have been considered very recently.
%
%
%
%

In the following, we first demonstrate the emergence of a proximity-induced PCC
by means of a band structure analysis within a simplified model calculation.
Subsequently, we will present numerical results that show the same features 
in  HgTe/CdTe-based ribbons.
Furthermore, we will study features of the PCC in quantum transport for 
experimentally realizable setups.
In particular we reveal the creation of a PCC in an otherwise
insulating disordered strip and how to disentangle the signature of a PCC from
the diffusive modes in a disordered conductor.

Throughout this manuscript we model the electronic structure of a 2d 
HgTe/CdTe quantum-well by the four-band Bernevig-Hughes-Zhang (BHZ)
Hamiltonian~\cite{V_Bernevig2006}, 
%
%
\begin{equation}
H=\left(\begin{array}{cc}
h(\mathbf{k}) & \begin{array}{cc}
0 & -\Delta\\
\Delta & 0\end{array}\\
\begin{array}{cc}
0 & \Delta\\
-\Delta & 0\end{array} & h^{*}(-\mathbf{k})\end{array}\right),
\label{eq:bhzhamilton}
\end{equation}
%
%
with spin-subblock Hamiltonians
%
%
\begin{equation}
h(\mathbf{k})=\left(\begin{array}{cc}
M-(B+D)\mathbf{k}^{2} & Ak_{+}\\
Ak_{-} & -M+(B-D)\mathbf{k}^{2}\end{array}\right),
\end{equation}
%
%
where $k_{\pm}=k_{x}\pm i k_{y}$ and $\mathbf{k}^{2}=k_{x}^{2}+k_{y}^{2}$.
In Eq.~\eqref{eq:bhzhamilton}, the off-diagonal coupling parameter $\Delta$ describes a
Dresselhaus-type spin-orbit coupling of the two spin blocks \cite{Rothe2010}.

%
%
\begin{figure}[t]
\centering
\includegraphics[width=\stdwidth]{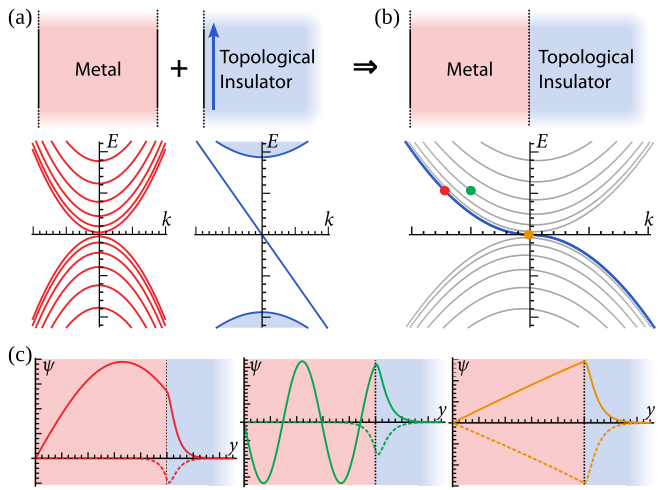}
\caption{ (\emph{Color online})
 Band structure of (\sub{a})~a non-topological strip (left) and a semi-infinite topological insulator plane (right), 
(\sub{b})~the combined system.
(\sub{c})~Wave functions for parameters marked by dots in (\sub{b}). Solid/Dashed lines show the 
first/second component of the wave functions.
}
\label{figBands}
\end{figure}
%
%

We consider a 2d heterostructure of a usual conductor and a semi-infinite
topological insulator plane.
It is instructive to first look at a simplified model system which already
captures the important physical features.
The metal part~\cite{note2} is modeled by the Hamiltonian~\eref{eq:bhzhamilton} 
with $A=D=M=\Delta=0$, describing a free electron-hole gas.
Confined to a finite-width strip by hard wall boundary conditions,
this leads to an electron-hole symmetric parabolic band structure,
as shown in Fig.~\subref{figBands}{a}.
For the TI half plane, we use a minimal BHZ Hamiltonian ($D=\Delta=0$),
which splits up into two electron-hole symmetric spin blocks.
The band structure for one spin block shown in Fig.~\subref{figBands}{a}
exhibits the bulk band gap and a single edge state.
A coupling between the two systems is implemented by
replacing the central hard wall constraint by the continuity of 
the wave function $\Psi(y)$ and the flux $v_y \Psi(y)$ across the interface.
The resulting band structure, shown in Fig.~\subref{figBands}{b},
resembles the one of the metallic strip, but
contains an extra band (marked in blue) as a remainder 
of the topological edge state, which does not feature the linear momentum
dependence because of band repulsion due to hybridization.
The wave functions at various energies in Fig.~\subref{figBands}{c} reveal that {\em all} states
are predominantly localized in the metallic region and only 
small exponential tails remain in the TI part.
Thus, we do not find evidence for an induced non-trivial topology as reported
in Ref.~\onlinecite{xiao2013proximity}, from which one would expect a new interface state
forming on the outer hard wall boundary.
Instead, the former edge state fully hybridizes with the metallic bands and covers the whole extended state region!
This is in line with Ref.~\onlinecite{Wang2013} where the authors also observe an extended
metallic state in a 3d heterostructure, but differs from the findings of Ref.~\onlinecite{Yang2013},
where a protection from hybridization for topological edge states is reported 
for a 2d quantum Hall insulator on top of a 3d substrate.

The above result has important consequences. 
Counting the bands in Fig.~\subref{figBands}{b} reveals an 
imbalance between left and right movers (up and down movers in Fig.~\ref{figBands}),
which leads to a PCC at any Fermi energy.
Including the other spin block and spin-orbit interaction 
only slightly changes this picture.
Since both spin blocks feature a PCC with opposite propagation direction in the decoupled system,
the combined system has an equal but {\em odd} number of left and right moving channels. 
Due to the time-reversal symmetry of the Hamiltonian,
the scattering matrix can be written in a basis where it is anti-symmetric.
Together with the odd number of channels, this implies a single PCC 
in each direction~\cite{Bardarson2008}.
As the wave functions of all channels are predominantly localized in the metallic region, one
can say that the proximity of the TI induces a single additional channel in each direction
leading to one net PCC in the metallic region.
In other words, the quasi 1d edge state of the TI migrates into the metallic
region and forms an extended PCC.
We believe that this is generally true for materials with extended states. It should
provide a valuable tool to create and observe PCCs as they inherit the topological
protection of the former edge channels and are expected to be stable for any disorder
and interface type, as long as the latter allows for sufficient hybridization.

Given that the TI is a good insulator, i.e., that the decay length of the
states inside the insulator is short compared to the width of the metallic ribbon, one
expects that only a vanishing part of the probability density of a state is
inside the TI.
This allows for replacing the explicit consideration of the TI by
an effective boundary condition for the wave function $\Psi(y)$ inside the metallic ribbon.
For the right boundary of a heterostructure in $y$-direction it reads:
%
%
\begin{eqnarray}
\Psi_{2} & = & -\Psi_{1}\label{eq:bcdiri}\\
i\left(v_{y}\Psi\right)_{2}-i\left(v_{y}\Psi\right)_{1} & = & 4 Q k_x\Psi_{1},
\label{eq:bcrobin}
\end{eqnarray}
%
%
where $\left(v_{y}\Psi\right)_{i}$ denotes the $i$-th component of the state obtained by applying the
velocity operator normal to the boundary.
Note that, Eq.~\eref{eq:bcrobin} still contain a free parameter $Q$, which stems from
the band curvature of the TI.
The effect $Q$ is best seen by applying the boundary conditions
to a topologically trivial insulator, like a gapped electron-hole gas:
Then the group velocity of the emergent edge state band linearly depends on $Q$.
Using the boundary conditions on the gapless electron-hole gas from the previous model heterostructure, Fig.~\subref{figBands}{a}, 
we obtain for $Q=B$ qualitatively the same band structure as in Fig.~\subref{figBands}{b}.
While Eq.~\eref{eq:bcdiri} is a simple Dirichlet boundary condition that fixes the phase of the
two components, Eq.~\eref{eq:bcrobin} is of the \emph{Robin type} \cite{gustafson1998third}, i.e., it links the wave function
at the boundary to its derivative.
This $k_x$-dependent mixing induces the PCC in the adjacent metal, as
illustrated in the Supplemental Material~\cite{support}.

\begin{figure}[t]
\centering
\includegraphics[width=\stdwidth]{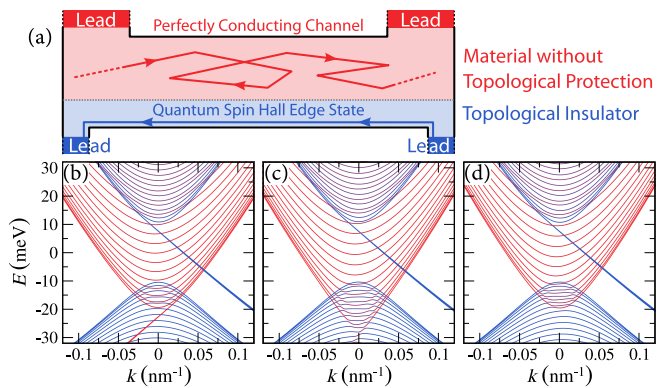}
\caption{ (\emph{Color online})
(\sub{a}) Scheme of the four-lead geometry used for PCC observation
(\sub{b})--(\sub{d}) Band structures of a nanoribbon ($W=350\,{\rm nm}$, HgTe material parameters \cite{support}) with 
different band topology [(\sub{b}) $M=-10\,{\rm meV}$, (\sub{c}) $M=0\,{\rm meV}$, (\sub{d}) $M=10\,{\rm meV}$]
in proximity to a TI ribbon (energy offset $E_0=30\,{\rm meV}$).
Colors encode the position of the corresponding wave function 
(red: metallic part, blue: TI).
}
\label{figBS}
\end{figure}

For the practical observation of a PCC in such heterostructures the semi-infinite TI plane
can be replaced by a TI-ribbon in a four-lead geometry, see Fig.~\subref{figBS}{a}.
This geometry decouples the edge channel on the lower edge of the TI from the
region with the PCC.
This is necessary, as the lower edge state would add another channel to the scattering matrix,
make it even dimensional and therefore lift the protection of the PCC.
If the PCC in the upper metal is well coupled to the two metallic leads (e.g., by making
the leads wide), the total transmission $T_\textrm{M}$ between the upper leads will be at least one.
Even in presence of disorder there will be no complete localization independent of the system length.

%
%
%
%
It is not required to assemble the heterostructure presented in Fig.~\subref{figBS}{a} 
from two different materials.
For convenience the whole system can consist of a common HgTe/CdTe quantum well
with inverted band order, where only the upper (metallic) part has a Fermi energy outside
the TI bandgap, a setting that can be realized by local gating.
Such a setup leads to a hybridization of the upper pair of edge states of the TI with the
bulk states of the metallic part, while the lower pair is still localized at the lower
edge of the TI.
The corresponding band structure in Fig.~\subref{figBS}{b} reveals that a single spin block
contains an additional right moving state (red lines), whereas the left mover is still
localized at the opposite boundary in the TI region (blue).
Furthermore, we stress that an inverted band order is not strictly necessary to 
induce a PCC.
The bands of a gapless metallic strip [see Fig.~\subref{figBS}{c}]
as well as a strip with conventional ordering [see Fig.~\subref{figBS}{d}] also feature an
additional right moving state in proximity to a TI.
This implies that PCCs can be induced by the same mechanism in other semiconductors if
the two materials can be coupled efficiently and the crossover of the band
topology takes place in the metallic part or at the interface.

\begin{figure}[t]
\centering
\includegraphics[width=\stdwidth]{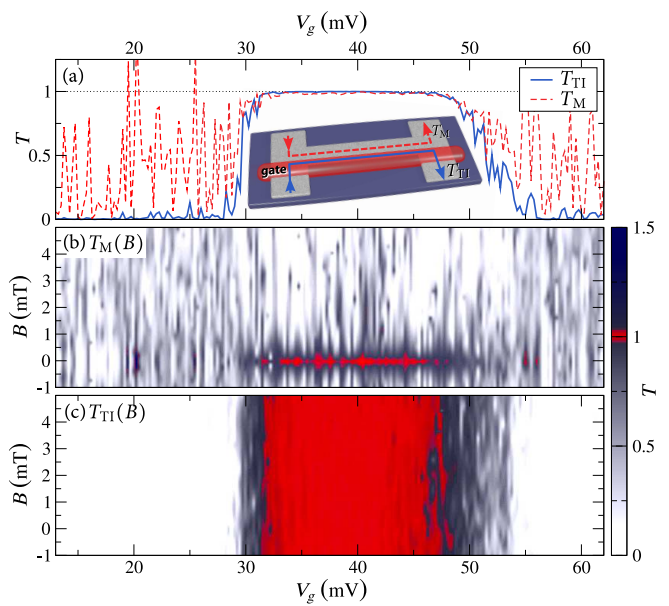}
\caption{ (\emph{Color online})
Quantized conductance of a perfectly conducting channel in a strongly disordered HgTe/CdTe ``H-bar"
($L\,{=}\,8\,{\rm \mu m}$, $W_{M}\,{=}\,350\,{\rm nm}$, $W_\mathrm{TI}\,{=}\,150\,{\rm nm}$, $W_L\,{=}\,800\,{\rm nm}$
as shown in the inset and ~\cite{support}).
(\sub{a})~Transmission $T_\textrm{M}$ (red dashed) between the upper leads and $T_\mathrm{TI}$ (blue solid) between the lower leads
featuring a quantized value for $V_g\in[30,50]\,{\rm mV}$ where only the lower gated part is in a TI state.
(\sub{b})~Magnetic field dependence of transmission $T_\textrm{M}(B)$ (color coded) through the ungated normal part.
(\sub{c})~Transmission $T_\mathrm{TI}(B)$ through the gated (TI) part.
}
\label{figHbar}
\end{figure}

Experimental evidence for the induction of a PCC can be
for example accomplished by means of magneto transport measurements
in the localized regime.
To this end, we suggest a four-terminal configuration based on a 2d-TI like
HgTe/CdTe, as sketched in the inset of Fig.~\subref{figHbar}{a}.
%
%
%
%
We assume the Fermi energy outside the topological band gap, such that propagating states
exist in the bulk of the corresponding clean material. 
For strong disorder associated with a localization length shorter than the length~$L$ of
the ribbon connecting the left and right terminals, the total transmission $T_\textrm{M}$ between the upper leads 
and $T_\mathrm{TI}$ between the lower leads is strongly suppressed.
To achieve TI properties in the lower part of the strip, a gate
is applied to locally shift the band structure of a region containing the edge and the entire
connection between the strip and the lower leads
[red region in the inset of Fig.~\subref{figHbar}{a}].
To illustrate the signatures of the PCC the transport properties
of this setup are calculated for different gate potentials $V_g$
by means of a recursive Greens function algorithm~\cite{Wimmer2009}.
If the gated part is tuned such that the Fermi energy is in
the TI bandgap, the suppressed transmission $T_\mathrm{TI}$ of the localized regime
takes the quantized value $T_\mathrm{TI}\approx 1$ of the quantum spin Hall edge state.
This behavior of $T_\mathrm{TI}$ is shown in Fig.~\subref{figHbar}{a} by the solid (blue)
line for gate voltages between $30\,\text{mV}$ and $50\,\text{mV}$.
Remarkably, the transmission  $T_\textrm{M}$ between the two upper leads exhibits the
same behavior and gets quantized, although there is no material with topological protection
linking the two terminals.
This quantized transmission in the otherwise localized regime is a clear-cut signature of and
carried by the PCC.

The different nature of the PCC compared to the quantum spin
Hall edge state can be probed by a perpendicular magnetic field $B$.
Since the PCC state is extended over the entire spatial region $A$ of the upper strip,
which can be seen in the local density of states~\cite{support},
a weak field generating a flux $A\cdot B\approx \phi_0$, where $\phi_0$ is the flux quantum, 
is sufficient to effectively break the time-reversal symmetry.
The numerical data presented in Fig.~\subref{figHbar}{b} shows that the conductance quantization
of $T_\textrm{M}$ is fully destroyed for a magnetic field of $B=2~\mathrm{mT}$,
corresponding to $\phi_0$ penetrating the device.
This is not the case for the robust quantum spin Hall edge state, which is strongly localized
at the lower boundary and accordingly quasi one-dimensional. 
As a result, the magneto transport is insensitive to magnetic fields 
in the $\mathrm{mT}$ regime, as shown for $T_{\textrm{TI}}$ in Fig.~\subref{figHbar}{c}.
As a side remark, the PCC turns out to be completely spin mixed already at small spin-orbit strengths $Delta$,
 in contrast to the quantum spin Hall edge channels, which remain spin-polarized up to moderate
spin-orbit strengths~\cite{support}.

If the electronic states of the disordered strip are not localized the total transmission
$T_\textrm{M} = \sum_n t_n$ also contains
contributions $0 < t_n < 1$ of non-perfectly conducting channels.
%
%
\begin{figure}[t]
\centering
\includegraphics[width=\stdwidth]{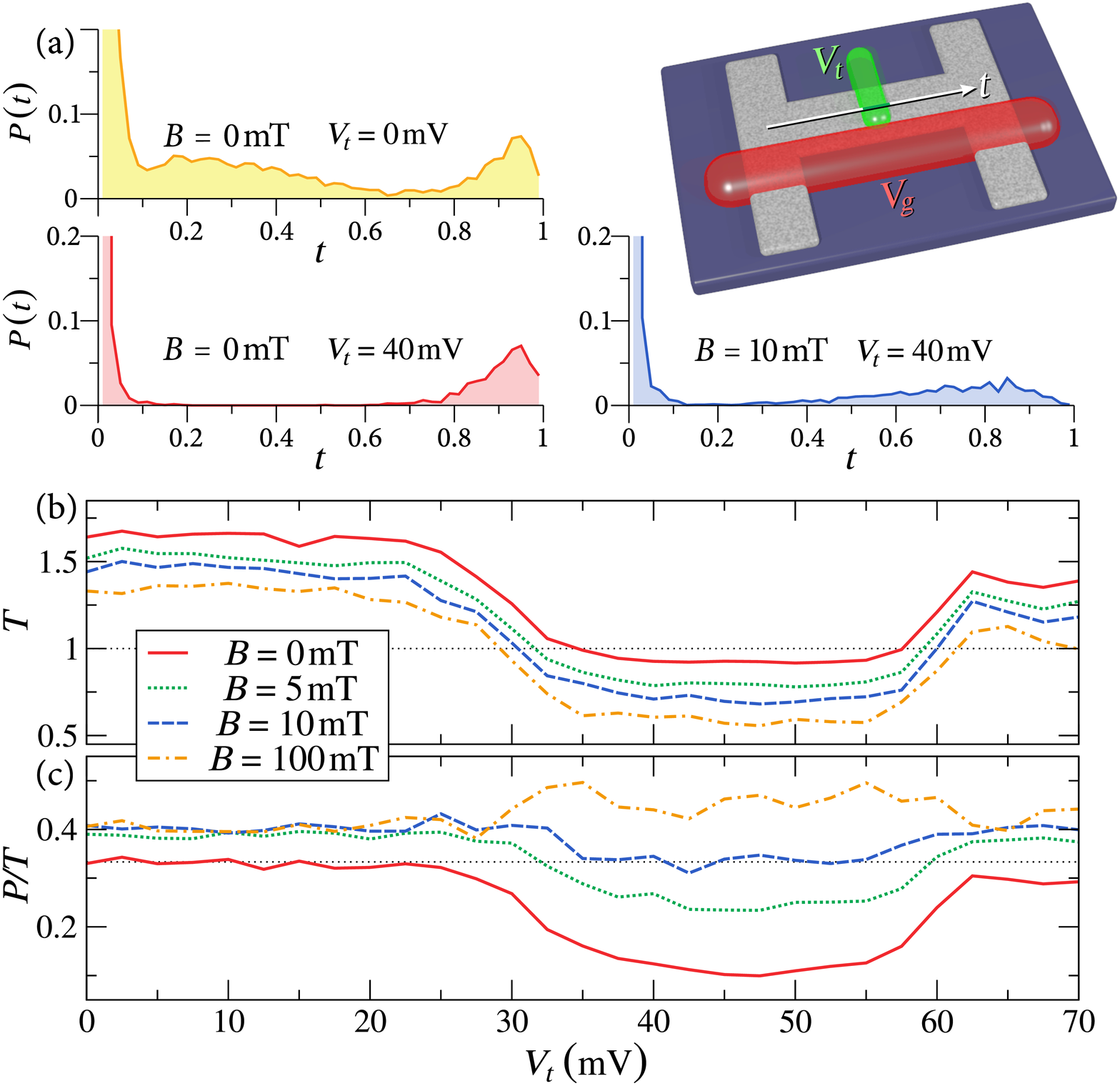}
\caption{ (\emph{Color online})
Tunnel setup for the detection of the perfectly conducting channel in the diffusive regime using an ``H-bar"
($L\,{=}\,1\,{\rm \mu m}$, $W_{M}\,{=}\,350\,{\rm nm}$, $W_\mathrm{TI}\,{=}\,150\,{\rm nm}$, $W_L\,{=}\,400\,{\rm nm}$)
and an additional tunnel gate $V_t$ as shown as green bar in the inset.
(\sub{a})~Distribution functions of the transmission eigenvalues $t$ (from ensemble averages over $1000$ impurity configurations) 
for different tunnel gates $V_t$ with and without magnetic field.
(\sub{b})~Average total transmission $T$ between the upper leads with $T\approx 1$ for $B=0$ in presence of a sufficiently strong tunnel barrier.
(\sub{c})~Average Fano factor $P/T$ for the current between the upper leads featuring a strong suppression of shot noise in presence of a
perfectly conducting channel. 
}
\label{figtunnel}
\end{figure}
%
%
For diffusive transport the distribution $P(t)$ of these transmission eigenvalues $t$ features a bimodal
distribution~\cite{Dorokhov1984381, RevModPhys.69.731}, see upper left panel of Fig.~\subref{figtunnel}{a},
whose intermediate transmission eigenvalues mask the PCC when considering $T_\textrm{M}$.
In this case an additional tunnel barrier, depicted by the green bar in the sketch of Fig.~\ref{figtunnel}, 
can be used to unravel the presence of the PCC.
The probability for finding these intermediate transmission eigenvalues can be strongly reduced by the tunnel
barrier (parametrized by $V_t$), as shown in Fig.~\subref{figtunnel}{a} for $B=0\,{\rm mT}$, $V_t=40\,{\rm mV}$.
Hence, for this configuration $T_\textrm{M}$ is mainly carried by the PCC.
As in the localized case, protection of the PCC can be broken by a magnetic flux
leading to a broader distribution for the high transmission eigenvalues
[right panel in Fig.~\subref{figtunnel}{a}].

The change of this probability distribution is also reflected in transport quantities, which can
be experimentally probed.
For example, upon raising the tunnel barrier $V_t$ the transmission at $B=0$ exhibits a quantized minimum,
which reflects the presence of a PCC,
as shown by the red line in Fig.~\subref{figtunnel}{b} for $V_t\in[30,60]\,{\rm mV}$. As expected, this
feature disappears at small magnetic fields.

A more pronounced PCC signature is found in the shot noise power~\cite{PhysRevLett.65.2901}, $P = \sum_n t_n(1-t_n)$,
depending on the transmission eigenvalues $t_n$ between the upper leads.
For conventional diffusive transport the bimodal distribution leads to a universal $1/3$
suppression of the shot noise [$P = T/3$, as marked by a dashed line in Fig.~\subref{figtunnel}{c}],
independent of the universality class of the material~\cite{PhysRevB.46.1889}.
In presence of a tunnel barrier all $t_n$ except the one of the PCC tend to zero,
leading to a characteristic shot noise suppression,
as shown by our numerics in Fig.~\subref{figtunnel}{c} for $V_t\in[30,60]\,{\rm mV}$.
A $B$-field destroys the PCC,
which in turn removes the shot noise suppression. At finite $B$, in absence of the PCC
the shot noise signal even increases above $P=T/3$ due to the tunnel barrier.
We suggest the tunnel conductance and shot noise as  
promising, experimentally accessible observables to verify the 
TI proximity induced PCC.

To summarize, we showed that the proximity of a 2d topological insulator (TI) creates a robust
perfectly conducting channel (PCC) in an adjacent non-topological material.
We believe this is a quite general effect that should work for almost all materials with
extended states. We expect that a similar phenomenon exists for a disordered conductor
at the surface of a 3d TI.
In addition, we showed that the proximity of a 2d TI to a metal can be understood in terms
of a Robin type effective boundary condition, responsible for the creation of the PCC.
This does not only allow simplifications in future theoretical studies, but might also pave the way
to induce PCCs by artificially creating such boundary conditions, e.g., by using
metamaterials for electromagnetic waves, without relying on TI heterostructures.

This work is supported by DFG
(SPP 1666 and joined DFG-JST Research Unit FOR 1483).
We thank I. Adagideli and M. Wimmer for useful conversations.

\vspace*{1cm}

\begin{appendix}

\section{Band structure calculations}
\label{sec::bands}

This section provides some details on the band structure calculations
shown in Fig.~2 of the main manuscript. We consider a metallic strip
of finite width $W$, which borders vacuum on the one side and is
in contact with a semi-infinite 2d-TI plane on the other. In the TI
part $H_{\text{TI}}$ is given by a simplified 2-band BHZ-model
\begin{equation}
H_{\text{TI}}=\left(\begin{array}{cc}
M-Bk^{2} & A\left(k_{x}+ik_{y}\right)\\
A\left(k_{x}-ik_{y}\right) & -M+Bk^{2}
\end{array}\right),
\end{equation}
where in the metallic region $H_{\text{met}}$ is simply taken as
a free electron-hole gas ($A=M=0$)
\begin{equation}
H_{\text{met}}=\left(\begin{array}{cc}
-Bk^{2} & 0\\
0 & Bk^{2}
\end{array}\right).\label{eq:free2comphamil}
\end{equation}
For simplicity we assume the value of the parameter $B$ to be the
same in the two regions. To solve for the transverse wave functions
in this geometry, we require that the wave function $\psi_{k_{x}}(y)$
should vanish at the boundary to vacuum at $y=0$ and that $\psi_{k_{x}}(y)$
should be continuous across the metal-TI boundary at $y=W$. We also
require the probability current to be continuous across the boundary,
i.e.,
\begin{eqnarray}
\left(\begin{smallmatrix}
2iB\partial_{y} & 0\\
0 & -2iB\partial_{y}
\end{smallmatrix}\right)\psi_{k_{x}}(W-\epsilon)=\left(\begin{smallmatrix}
2iB\partial_{y} & iA\\
-iA & -2iB\partial_{y}
\end{smallmatrix}\right)\psi_{k_{x}}(W+\epsilon).
\end{eqnarray}
We then find the secular equation
\begin{widetext}
\begin{eqnarray}
 &  & \frac{4B^{2}f_{p}\left(\lambda_{2}u_{1}v_{2}-\lambda_{1}u_{2}v_{1}\right)}{f_{m}\cosh(f_{m}W)\sinh(f_{p}W)}
 +\frac{\left(4B^{2}-A^{2}\right)\lambda_{1}\lambda_{2}\left(u_{2}v_{1}-u_{1}v_{2}\right)-2AB\left(\lambda_{1}-\lambda_{2}\right)\left(u_{1}u_{2}-v_{1}v_{2}\right)}{f_{m}f_{p}\cosh(f_{m}W)\cosh(f_{p}W)}\\
 & = & \frac{4B^{2}f_{p}f_{m}\left(u_{1}v_{2}-u_{2}v_{1}\right)}{\sinh(f_{m}W)\sinh(f_{p}W)}+\frac{4B^{2}f_{m}\left(\lambda_{2}u_{2}v_{1}-\lambda_{1}u_{1}v_{2}\right)}{\sinh(f_{m}W)\cosh(f_{p}W)},\label{eq:secequ}
\end{eqnarray}
\end{widetext}
which may be solved numerically to yield the band structure. Here,
\begin{eqnarray}
f_{p} & = & \sqrt{k_{x}^{2}+\frac{E}{B}},\\
f_{m} & = & \sqrt{k_{x}^{2}-\frac{E}{B}},
\end{eqnarray}
and $u_{1/2},v_{1/2}$ are the components of the decaying free solutions
$\xi_{k_{x}}^{1/2}(y)$ in the TI-region,
\begin{equation}
\xi_{k_{x}}^{1/2}(y)=\left(\begin{array}{c}
u_{1/2}\\
v_{1/2}
\end{array}\right)e^{\lambda_{1/2}y},
\end{equation}
\begin{eqnarray}
u_{1/2} & = & -\frac{E+B\left(\lambda_{1/2}^{2}-k_{x}^{2}\right)+M}{A\left(\lambda_{1/2}^{2}-k_{x}\lambda_{1/2}\right)},\\
v_{1/2} & = & \frac{1}{\lambda_{1/2}},
\end{eqnarray}
where $H_{\text{TI}}\xi_{k_{x}}^{1/2}(y)=E\xi_{k_{x}}^{1/2}(y)$.
The decay coefficients (which have a negative real part as long as
$\left|E\right|<\left|M\right|$) are given by
\begin{equation*}
\lambda_{1,2}  =  -\sqrt{\tfrac{A^{2}+2B^{2}k_{x}^{2}-2BM\mp\sqrt{A^{4}+4B^{2}E^{2}-4A^{2}BM}}{2B^{2}}}.
\end{equation*}
Plots of a typical bandstructure obtained from this model are shown
in Fig.~2(b) of the main text together with some sample wavefunctions
in Fig.~2(c). They were obtained using $A=5,\, B=-100,\, M=-0.05,\, W=1000$.

\section{Effective boundary condition}
\label{sec::boundary}

\subsection{Derivation\label{sub:Derivation}}

One can increase the level of abstraction and still observe the same
physics, by not explicitly considering the semi-infinite 2d TI plane
but instead replacing it by an effective boundary condition. This
is especially justified if the extent of the evanescent wave component
into the 2d-TI is small ($\frac{1}{\left|\textrm{Re}\lambda_{1/2}\right|}\ll1$),
such that the transverse wave function is to a good approximation
only located inside the metallic region. As the considered model of
the TI involves many parameters, there is no unique way of achieving
this infinitely fast decay and hence there will be no unique set of
boundary conditions. The limiting procedure presented in the following
should be understood as one possible way, which will in the end provide
a conceptually simple boundary condition, which will capture all important
physical features.

For achieving the infinite decay constants we let $M\rightarrow-\infty$,
i.e., we will consider a TI with an infinite band gap. In addition,
we will have to rescale $A$ by choosing $A=X\cdot\sqrt{-M}$ with
a parameter $X$ having the units $\sqrt{\textrm{energy}}\cdot\textrm{length}$.
If one now considers the limit $M\rightarrow-\infty$ keeping $X=\frac{A}{\sqrt{-M}}=\textrm{const}$, one finds for
the real parts of the decay constants $\lambda_{1/2}$:
\begin{eqnarray}
\lim_{M\rightarrow-\infty}\textrm{Re}\,\lambda_{1} & = & -\infty,\\
\lim_{M\rightarrow-\infty}\textrm{Re}\,\lambda_{2} & = & -\infty,
\end{eqnarray}
meaning that we indeed expect the wave functions not to extend into
the TI. For this limit, one can derive the appropriate boundary conditions
that any attached two-band Hamiltonian should fulfill. We start from
the continuity of the wave function of the metal region $\psi^{M}(y)$
and the probability current $v_{y}\psi^{M}(y)$ across the boundary
at $y=W$,
\begin{eqnarray}
\left(\begin{array}{c}
\psi_{1}^{M}(W)\\
\psi_{2}^{M}(W)
\end{array}\right) & = & c_{1}\left(\begin{array}{c}
u_{1}\\
v_{1}
\end{array}\right)+c_{2}\left(\begin{array}{c}
u_{2}\\
v_{2}
\end{array}\right)\label{eq:wf_cont_1}\\
\left(-i\right)v_{y}\left(\begin{array}{c}
\psi_{1}^{M}(W)\\
\psi_{2}^{M}(W)
\end{array}\right) & = & c_{1}2B\lambda_{1}\left(\begin{array}{c}
u_{1}\\
-v_{1}
\end{array}\right)+c_{2}2B\lambda_{2}\left(\begin{array}{c}
u_{2}\\
-v_{2}
\end{array}\right)\nonumber \\
 &  & +Ac_{1}\left(\begin{array}{c}
v_{1}\\
-u_{1}
\end{array}\right)+Ac_{2}\left(\begin{array}{c}
v_{2}\\
-u_{2}
\end{array}\right)\label{eq:wfp_cont_1}
\end{eqnarray}
for so far unknown constants $c_{1}$ and $c_{2}$. We are now interested
in the behavior of the right hand sides for the limit discussed
above. However, as
\begin{equation}
\lim_{M\rightarrow-\infty}\left|u_{1/2}\right|=\lim_{M\rightarrow-\infty}\left|v_{1/2}\right|=0,
\end{equation}
one needs to rescale the constants to obtain a finite value of the
wave function at the border to the TI:
\begin{equation}
c_{1/2}^{\prime} = \frac{c_{1/2}}{\sqrt{-M}}.
\end{equation}
This way, we find
\begin{eqnarray}
\psi_{1}^{M}(W) & = & c_{1}^{\prime}\sqrt{-M}u_{1}+c_{2}^{\prime}\sqrt{-M}u_{2}\\
\psi_{2}^{M}(W) & = & c_{1}^{\prime}\sqrt{-M}v_{1}+c_{2}^{\prime}\sqrt{-M}v_{2},
\end{eqnarray}
from Eq.~\eqref{eq:wf_cont_1}. We can now do the limiting process
on the right hand side always keeping $X=\frac{A}{\sqrt{-M}}$ fixed, using
\begin{eqnarray}
\lim_{M\rightarrow-\infty}u_{1}\sqrt{-M} & = & \underbrace{\frac{1}{2}\left(X+\sqrt{4B+X^{2}}\right)}_{\equiv q_{1}(B,X)}\\
\lim_{M\rightarrow-\infty}v_{1}\sqrt{-M} & = & -\underbrace{\frac{1}{2}\left(X+\sqrt{4B+X^{2}}\right)}_{\equiv q_{1}(B,X)}\\
\lim_{M\rightarrow-\infty}u_{2}\sqrt{-M} & = & \underbrace{\frac{1}{2}\left(X-\sqrt{4B+X^{2}}\right)}_{\equiv q_{2}(B,X)}\\
\lim_{M\rightarrow-\infty}v_{2}\sqrt{-M} & = & -\underbrace{\frac{1}{2}\left(X-\sqrt{4B+X^{2}}\right)}_{\equiv q_{2}(B,X)}
\end{eqnarray}
where, on the way, we made use of the fact that $B<0$ and $X>0$.
To simplify the writing we introduced the functions $q_{1}(B,X)$
and $q_{2}(B,X)$. From this, we can already find the first boundary
condition, as now
\begin{eqnarray}
\psi_{1}^{M}(W) & = & c_{1}^{\prime}q_{1}(B,X)+c_{2}^{\prime}q_{2}(B,X),\\
\psi_{2}^{M}(W) & = & -c_{1}^{\prime}q_{1}(B,X)-c_{2}^{\prime}q_{2}(B,X),\\
\rightarrow\psi_{1}^{M}(W) & = & -\psi_{2}^{M}(W).
\end{eqnarray}
Note that this is independent of the choice of the parameters $B$
and $X$. The second boundary condition which links the currents is,
however, more complicated and does depend on the choice of $B$ and
$X$. It can be simplified by additionally choosing $X=2\sqrt{-B}$,
which will then imply
\begin{equation}
q_{1}\left(B,X=2\sqrt{-B}\right)=q_{2}\left(B,X=2\sqrt{-B}\right)=\sqrt{-B}.
\end{equation}
If we now subtract the two components of Eq.~\eqref{eq:wfp_cont_1},
\begin{eqnarray}
 &  & i\left(v_{y}\psi^{M}\right)_{2}(W)-i\left(v_{y}\psi^{M}\right)_{1}(W)\nonumber \\
 & = & c_{1}^{\prime}\left(2B\lambda_{1}+A\right)\left(u_{1}+v_{1}\right)\sqrt{-M}\nonumber \\
 &   & +c_{2}^{\prime}\left(2B\lambda_{2}+A\right)\left(u_{2}+v_{2}\right)\sqrt{-M}.
\end{eqnarray}
and again perform the limits,
\begin{eqnarray}
 \lim_{M\rightarrow-\infty}\left(2B\lambda_{1}+A\right)\left(u_{1}+v_{1}\right)\sqrt{-M} & = & 4B\sqrt{-B}k_{x}\\
 \lim_{M\rightarrow-\infty}\left(2B\lambda_{2}+A\right)\left(u_{2}+v_{2}\right)\sqrt{-M} & = & 4B\sqrt{-B}k_{x},
\end{eqnarray}
we obtain
\begin{eqnarray}
i\left(v_{y}\psi^{M}\right)_{2}(W)-i\left(v_{y}\psi^{M}\right)_{1}(W) & = & 4Bk_{x}\left(c_{1}^{\prime}\sqrt{-B}+c_{2}^{\prime}\sqrt{-B}\right)\nonumber \\
 & = & 4Bk_{x}\psi_{1}^{M}(W) \\
 & = & 2Bk_{x}\left(\psi_{1}^{M}(W)-\psi_{2}^{M}(W)\right).\nonumber
\end{eqnarray}
Here, we already made use of the first boundary condition. In total
we find, that the boundary to a TI in the limit discussed above can
be approximately described by the following boundary conditions, which
still contain one free parameter,
\begin{eqnarray}
\psi_{1}(W) & = & -\psi_{2}(W)\label{eq:bcdiricl}\\
i\left(v_{y}\psi\right)_{1}(W)-i\left(v_{y}\psi\right)_{2}(W) & = & 2Qk_{x}\left(\psi_{1}(W)-\psi_{2}(W)\right),\label{eq:bcrobin}
\end{eqnarray}

\noindent which we renamed from $B$ to $Q$, to distinguish it from the band
curvature of the material on which we enforce this boundary condition.
Like this we can generalize the case discussed in Sec.~\ref{sub:Derivation}
and need not restrict ourselves to heterostructures of materials with
the same band curvature. 
\begin{figure}[t]
\centering
\includegraphics[width=\stdwidth]{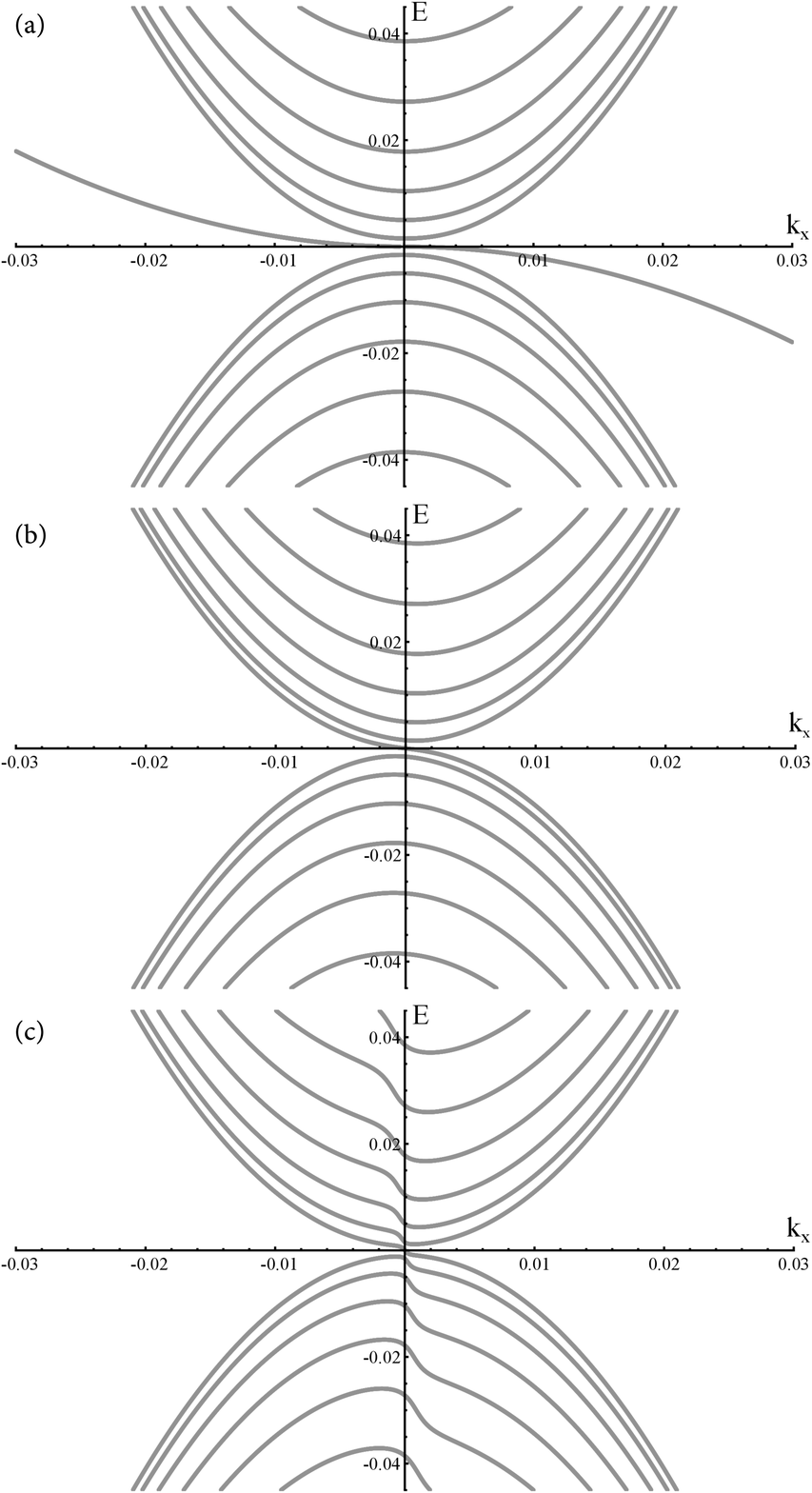}
\caption{\label{fig:bcwithdifferentqs}Band structures calculated with a free
electron-hole gas Hamiltonian, Eq.~\eqref{eq:free2comphamil}, using
band curvature $B=-100$, subject to hard wall boundary conditions
at $y=0$ and the boundary conditions from Eqs.~\eqref{eq:bcdiricl}
and \eqref{eq:bcrobin} at $y=1000$. The plots differ in the choice
of the parameter $Q$: (a) $Q=-10$, (b) $Q=B=-100$, (c) $Q=-1000$.}
\end{figure}
Fig.~\eqref{fig:bcwithdifferentqs} shows band structure calculations
using the boundary conditions with different choices for the parameter
$Q$. One notes that for $Q=B$ one almost recovers the band structure
from Fig.~2(b) of the main text which was obtained by a full calculation
which explicitly includes the TI half-plane. 

In the derivation, we assumed the TI half plane to extend in the positive
$y$-direction, which is why we matched the exponentially decaying
wave functions on the TI side. Of course the same calculation can
in principle be redone for an arbitrary interface orientation. E.g.,
choosing the TI plane to extend in the negative $y$-direction with
a boundary at $y=0$, leads to the following set of boundary conditions:

\begin{eqnarray}
\psi_{1}(0) & = & \psi_{2}(0)\\
i\left(v_{y}\psi\right)_{1}(0)+i\left(v_{y}\psi\right)_{2}(0) & = & -2Bk_{x}\left(\psi_{1}(0)+\psi_{2}(0)\right),\label{eq:bcrobin2}
\end{eqnarray}
i.e., we find an additional phase factor in front of one wave function
component.

\subsection{Effect of Robin boundary condition}

\begin{figure}
\begin{centering}
\includegraphics[width=6cm]{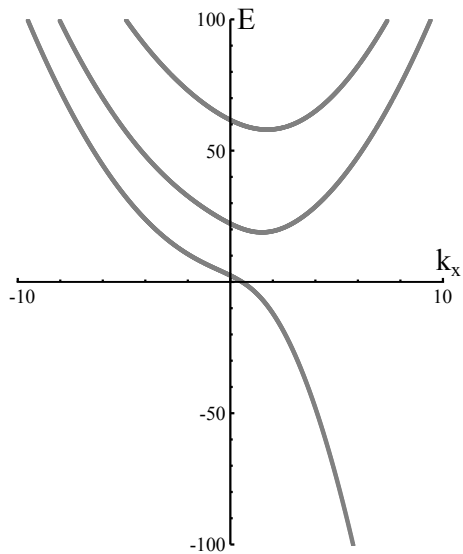}
\par\end{centering}

\caption{\label{fig:simple1componentrobin}Band structure of a wave guide with
a free electron gas Hamiltonian, Eq.~\eqref{eq:freehamilt}, subject
to hard wall boundary conditions at $y=0$, and the Robin boundary
condition from Eq.~\eqref{eq:bcrobin3} at $y=1$ using $\frac{1}{2}Q=B=1$.}
\end{figure}

To show that it is indeed the Robin boundary condition, Eq.~\eqref{eq:bcrobin}
or \eqref{eq:bcrobin2}, which is responsible for the appearance of
a perfectly conducting channel, we consider a one-component free electron
gas Hamiltonian 
\begin{equation}
H=B\left(k_{x}^{2}+k_{y}^{2}\right)\label{eq:freehamilt}
\end{equation}
 subject to hard wall boundary conditions at $y=0$ ($\psi(0)=0$)
and a one-component version of the boundary condition from Eq.~\eqref{eq:bcrobin}
at $y=W$:
\begin{eqnarray}
i\left(v_{y}\psi\right)(W)=2B\partial_{y}\psi(W) & = & -2Qk_{x}\psi(W)\\
\rightarrow\partial_{y}\psi(W) & = & -\frac{Q}{B}k_{x}\psi(W).\label{eq:bcrobin3}
\end{eqnarray}
For $\left|Q\right|>\left|B\right|$ one obtains a band structures
similar to the one shown in Fig.~\ref{fig:simple1componentrobin},
which has been calculated by setting $\frac{1}{2}Q=B=1$. It resembles
the simple parabolic band structure that one obtains by putting hard
wall (or Neumann) boundary conditions on both sides, but it contains
an extra band which tends to negative energies with increasing $k_{x}$
and thereby creates a PCC. While the emergence of large negative eigenvalues
is known for Robin problems $\partial_{y}\psi=-\alpha\psi$ with large
negative parameters $\alpha$ \cite{levitin2008principal}, as far
as we know there has not been a discussion on the emergence of perfecly
conducting channels in one-sided Robin wave guides so far.

By implementing this boundary condition, or a similar one (e.g. by replacing $k_{x}$ in Eq.~\eqref{eq:bcrobin3} by a different
but still odd function of $k_{x}$), on one side of a wave guide,
one should be able to create PCCs for many scenarios, i.e., for quantum
as well as electromagnetic wave guides.

\section{Transport Calculations}
\label{sec::transport}
In the manuscript we use numerical quantum transport calculations to investigate the
transport signatures of a PCC, which is induced in a conventional metal by the 
proximity of a TI.
To this end, we propose two different setups, which will be introduced in more detail in the following.
Both of them are based on a common HgTe/CdTe heterostructure, described by the 
BHZ Hamiltonian~(1) using the material parameters summarized in Table~\ref{materialparameters1}.
\begin{table}[b]
\centering
\begin{tabular}{c c c c c c c}
  \hline
  A & B & D & M & $\Delta$\\
  \hline
  $354.5$ & $-686$ & $-512$ & $-10$ & $1.6$\\ 
  \hline
\end{tabular}
\caption{\label{materialparameters1}Material parameters for HgTe/CdTe quantum wells (units in $\mathrm{meV}$ and $\mathrm{nm}$), taken from Ref.~\cite{Konig2008}.}
\end{table}

\begin{figure}[t]
\centering
\includegraphics[width=\stdwidth]{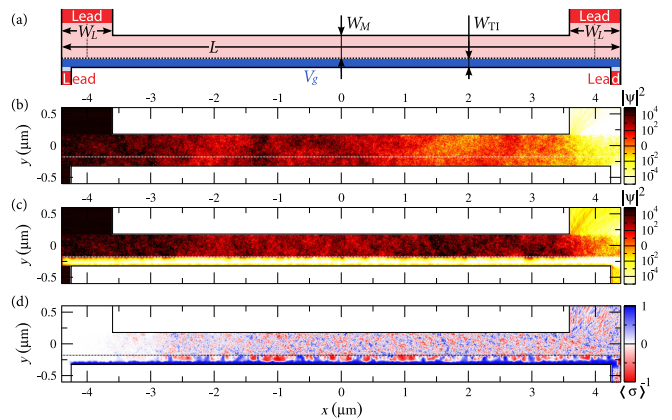}
\caption{ (\emph{Color online})
\vsub{a})
Sketch of the setup showing the perfectly conducting channel in transport through a strongly
disordered HgTe ``H-bar'' used in Fig.~4 of the main manuscript
($L\,{=}\,8\,{\rm \mu m}$, $W_M\,{=}\,350\,{\rm nm}$, $W_\mathrm{TI}\,{=}\,150\,{\rm nm}$, $W_L\,{=}\,800\,{\rm nm}$).
\vsub{b})~Typical carrier distribution (local density of states) arising for a small bias with higher chemical
potentials applied to the two left leads without additional gating
($E_F\,{=}\,-40\,{\rm meV}$, dark colors represent higher densities).
\vsub{c})~Typical carrier distribution for a gate potential of  $V_g\,{=}\,40\,{\rm mV}$.
\vsub{d})~Corresponding spin polarization. 
}
\label{figgeo}
\end{figure}

As first transport example, we simulate a H-shaped structure consisting of a long strip,
which is connected to four leads, as sketched in Fig.~\vsubref{figgeo}{a}.
%
%
%
%
%
The whole system consists of a single spatially constant material, which is discretized on a
square grid with a lattice spacing of $4~\mathrm{nm}$.
We assume the material hole doped and set the Fermi energy to $E_F\,{=}\,-40\,{\rm meV}$,
such that no quantum Hall edge states are involved in transport and the sample
behaves like a conventional metal.
Furthermore, Anderson disorder is added on each lattice site. 
The strength of this disorder is set to $U_0\,{=}\,30\,{\rm meV}$ and the length of the strip to $L\,{=}\,8\,{\rm \mu m}$,
which leads to a strong localization and a very low transmission between the right and the left side.
The signatures of the strong localization are visible in the non-equilibrium local density of states (LDOS)
in Fig.~\vsubref{figgeo}{b}, which is 
calculated for a potential gradient using higher chemical potentials for the two leads at the left side. 
The states entering the system at the left are not able to
propagate through the whole sample, which leads to a gradient in the LDOS from high densities 
(dark colors) at the left side to low densities (light colors) at the right side. 

In the following, a quantum spin Hall effect is induced by gating a part of the material into the TI
bandgap at the lower edge of the structure, as depicted by the blue region in Fig.~\vsubref{figgeo}{a}.
This results in a quantized conductance between the lower contacts, carried by a
quantum spin Hall edge state, which shows up as the thin dark connection
between the lower leads in the LDOS presented in Fig.~\vsubref{figgeo}{c}.
At the same time a PCC arises between the two upper leads and gives rise to quantized 
transport, as shown by the data in Fig.~4 of the manuscript.  
In comparison to the quantum spin Hall state this state is completely spread out over the whole upper part.
Furthermore, the spin polarization of the PCC differs strongly from the spin polarization of the edge state.
In the quantum spin Hall regime the edge state shows a very strong spin polarization,
visible by the constant blue color in the spin density at the lower edge in Fig.~\vsubref{figgeo}{d}.
On the contrary the PCC has no polarization, which can be seen by the patchy pattern 
in the upper part of the spin density.

%
%
\begin{figure}[t]
\centering
\includegraphics[width=\stdwidth]{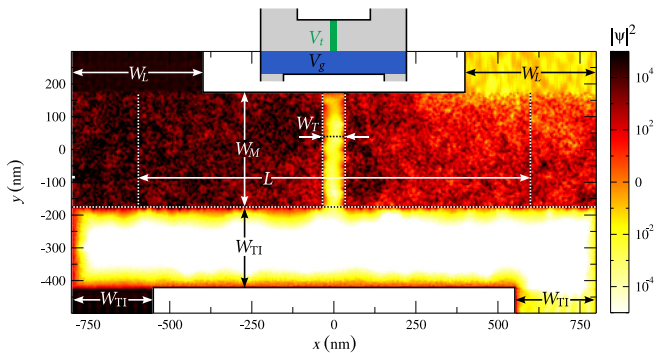}
\caption{ (\emph{Color online})
Tunnel setup for the detection of the perfectly conducting channel in the diffusive regime using an ``H-bar"
($L\,{=}\,1.2\,{\rm \mu m}$, $W_M\,{=}\,350\,{\rm nm}$, $W_\mathrm{TI}\,{=}\,250\,{\rm nm}$, $W_L\,{=}\,400\,{\rm nm}$, $W_T\,{=}\,120\,{\rm nm}$)
made of a HgTe/CdTe heterostructure. The same setup is used for the results presented in Fig.~5
of the main manuscript.
Typical carrier distribution arising for a small bias with higher chemical potentials applied to the left leads
($E_F\,{=}\,-40\,{\rm meV}$, dark colors represent higher densities).
Lower part is tuned into the topological bandgap by $V_g\,{=}\,40\,{\rm mV}$ and the upper strip is divided into two
parts by an additional tunnel barrier with $V_t\,{=}\,40\,{\rm mV}$.
}
\label{figtunnelgeo}
\end{figure}
%
%
In the previous setup a quantized conductance was used to demonstrate the emergence of a PCC
in an otherwise localized region.
Therefore a very long system was needed, such that all channels, except the PCC, are sufficiently 
localized by the impurity potential.
Although the PCC is still present if $L$ is smaller, its transport signature is masked by the
additional diffusive non-localized modes.
In this case an additional tunnel barrier can be used to suppress the influence 
of the non-perfectly conducting channels as shown in the manuscript.
Therefore, we investigate the transport properties in presence of such an additional tunnel barrier,
depicted by the green bar in the sketch of Fig.~\ref{figtunnelgeo}.
Similar to the previous setup, a gate $V_g$ is used to induce a quantum spin Hall state between the
lower leads and a PCC in the upper part, which can be seen in the LDOS in Fig.~\ref{figtunnelgeo}.
The tunnel barrier $V_t$ disconnects the upper two leads and reduces the contributions of the 
non-perfectly conducting channels.
As shown in Fig.~5 of the main manuscript, this leads to a vanishing shot noise and a quantized conductance.

\end{appendix}

\end{document}